\documentclass[graybox]{svmult}

\usepackage{mathptmx}       
\usepackage{helvet}         
\usepackage{courier}        
\usepackage{type1cm}        
\usepackage{makeidx}         
\usepackage{graphicx}        
\usepackage{multicol}        
\usepackage[bottom]{footmisc}

\usepackage{amsmath,amssymb}        

\makeindex             

\begin{document}

\title*{Efficient use of simultaneous multi-band observations for variable star analysis}
\titlerunning{Use of multi-band observations for variability analysis} 
\author{Maria S\"uveges, Paul Bartholdi, Andrew Becker, \v{Z}eljko Ivezi\'c, Mathias Beck and Laurent Eyer}
\authorrunning{M. S\"uveges, P. Bartholdi, A. Becker, \v{Z}. Ivezi\'c, M. Beck and L. Eyer}
\institute{Maria S\"uveges \at ISDC Data Centre for Astrophysics, Astronomical Observatory of Geneva, Ch. d'Ecogia 16, 1290 Versoix, Switzerland \email{Maria.Suveges@unige.ch}
\and Paul Bartholdi \at Astronomical Observatory of Geneva, Ch. des Maillettes 51, 1290 Sauverny, Switzerland \email{Paul.Bartholdi@unige.ch}
\and Andrew Becker \at University of Washington, 3910 15th Ave NE, Seattle WA 98195-1580, USA  \email{becker@astro.washington.edu}
\and \v{Z}eljko Ivezi\'c \at University of Washington, 3910 15th Ave NE, Seattle WA 98195-1580, USA  \email{ivezic@astro.washington.edu}
\and Mathias Beck \at ISDC Data Centre for Astrophysics, Astronomical Observatory of Geneva, Ch. d'Ecogia 16, 1290 Versoix, Switzerland \email{Mathias.Beck@unige.ch}
\and Laurent Eyer \at Astronomical Observatory of Geneva, Ch. des Maillettes 51, 1290 Sauverny, Switzerland \email{Laurent.Eyer@unige.ch}}
%
%
\maketitle

\abstract*{}

\abstract{The luminosity changes of most types of variable stars are correlated in the different wavelengths, and these correlations may be exploited for several purposes: for variability detection, for distinction of microvariability from noise, for period search or for classification. Principal component analysis is a simple and well-developed statistical tool to analyze correlated data. We will discuss its use on variable objects of Stripe 82 of the Sloan Digital Sky Survey, with the aim of identifying new RR Lyrae and SX Phoenicis-type candidates. The application is not straightforward because of different noise levels in the different bands, the presence of outliers that can be confused with real extreme observations, under- or overestimated errors and the dependence of errors on the magnitudes. These particularities require robust methods to be applied together with the principal component analysis. The results show that PCA is a valuable aid in variability analysis with multi-band data.
}

\section{Introduction}
\label{sec:intro}

In the recent era of large-scale astronomical surveys, the application of automated methods for the data processing and analysis has become indispensable. The tremendous amounts of data cannot be dealt with manually, as was done before. However, automatic procedures are rarely able to deal with exceptional, rare or radically new objects, and results have unavoidably higher error rates in the absence of direct interaction with the data. Thus, for characterizing and classifying the objects of a new survey, there is a need for better, more efficient extraction of information from the data that can improve the results of the automated procedures.

As the variations of different types of variable stars show distinct correlated color and luminosity light curve patterns, a promising possibility for variability detection, characterization and classification is the use of colors. A number of sky surveys produce multi-band data, like 
the Gaia satellite \cite{gaiaweb} or SDSS \cite{sdssweb} among others. We apply a well-known fundamental methodology of statistics, the principal component analysis (PCA) as a tool to combine quasi-simultaneous multi-filter observations, in order to obtain a reliable variability criterion based on  the presence of correlated variations, improved precision in period search and new features for variable type classification. The application must deal with several difficulties: astronomical observations are often encumbered  by under- or overestimated errors, band-wise sharply different error levels, outliers and other impediments. We use variance stabilizing transformation to reduce all bands to have equal error level, then a robust variant of PCA to linearly combine the simultaneous observations into the time series of PC1. This exhibits better signal-to-noise ratio than the single bands, and is therefore excellent for period search. The estimated parameters of the robust PCA fit on the scaled data can be used in various ways: the variance of the first principal component (PC1) or its ratio to the total variance measures the coherent variations at different wavelengths and therefore suggests intrinsic variability, while the coefficients in the linear combination PC1 are characteristic to the geometric or pulsational origin of the variability. These possibilities are tested on SDSS Stripe 82 unresolved sources as tools in variability analysis, in order to select new RR Lyrae and SX Phoenicis candidates, which are both Population II objects and are appropriate to trace old Galactic structures.


\section{Data}
\label{sec:data}

The Sloan Digital Sky Survey database provides five-band ($u$, $g$, $r$, $i$ and $z$) photometry of around 7500 deg$^2$ in the northern Galactic cap and around 740 deg$^2$ in the southern. One of the southern stripes, Stripe 82 was observed repeatedly during the first phase SDSS-I and the following SDSS-II Supernova Survey \cite{sesaretal07,bramichetal08,friemanetal08}, resulting in time series consisting of 30 observations on average for objects brighter than 21 mag in $g$, with photometric precision of 0.02 mag at the bright end and around 0.05 mag at the faint end. 

Our data set is a part of the catalog of \cite{sesaretal10}, separated into around 68,000 variable and 200,000 nonvariable objects based on rms scatter and chi-squared statistics cuts measured on $g$ and $r$ bands, as described in \cite{sesaretal07,sesaretal10}. Moreover, \cite{sesaretal10} presents a selection of confirmed RR Lyrae variables, on which we test our new procedures and which serves as a training set for our selection of additional RR Lyrae variables.

\section{Principal Component Analysis}
\label{sec:pca}

\begin{figure}[h]
\sidecaption
\includegraphics[scale=.5]{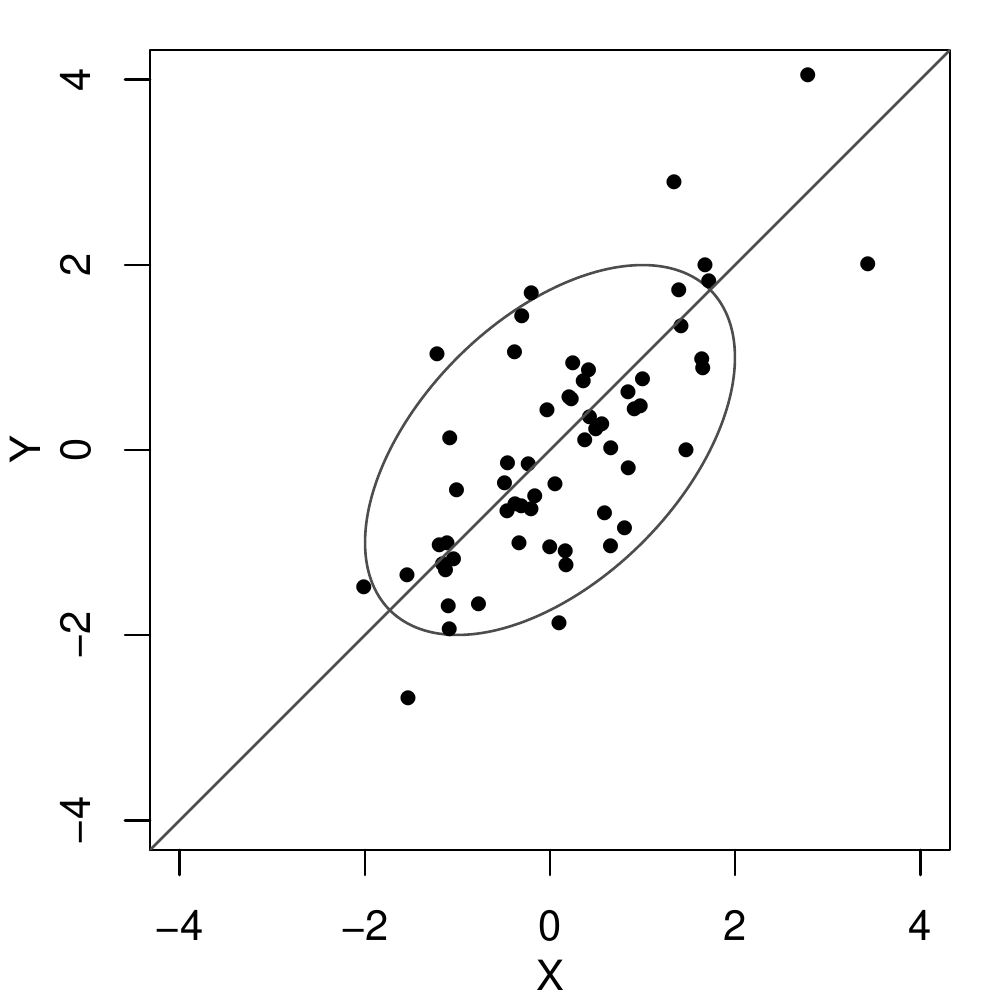}
\caption{Simulated standard normal variables with correlation 0.5. The direction of the largest scatter, the first principal direction is plotted as a line, together with an ellipsoid of equal density values of the 2-variate normal distribution.}
\label{fig:pca}       
\end{figure}

Suppose we have $N$ vector-valued data points, visualized as a point cloud in an $M$-dimensional space. Principal component analysis \cite{jolliffe02,hastieetal09} finds first the direction to which the projections of the points have the largest empirical variance;  then in the subspace orthogonal to this direction, repeats the operation iteratively, until it finds $M$ successive orthogonal directions. Mathematically, this is equivalent to finding the eigendecomposition of the empirical variance-covariance matrix of the data; the diagonal matrix of this decomposition is the variance-covariance matrix of the projections of the points to the found directions (called principal components). The direction of the maximal variance, termed PC1, is illustrated in Fig. \ref{fig:pca} for two correlated random variables, showing that the direction of PC1 is related to the presence of the correlation. In principle, if this point cloud represents the observed magnitudes of a variable star, the time series of the PC1 will likely give better period search results, since it has higher amplitude than any of the original variables, and in case of equal errors in the two bands, there is a $\sqrt{2}$ gain in the signal-to-noise ratio. When the point cloud corresponds to a nonvariable star, we do not expect correlation between the two bands, and we should observe a ball-like shape.

\begin{figure}[t]
\sidecaption[t]
\includegraphics[scale=.64]{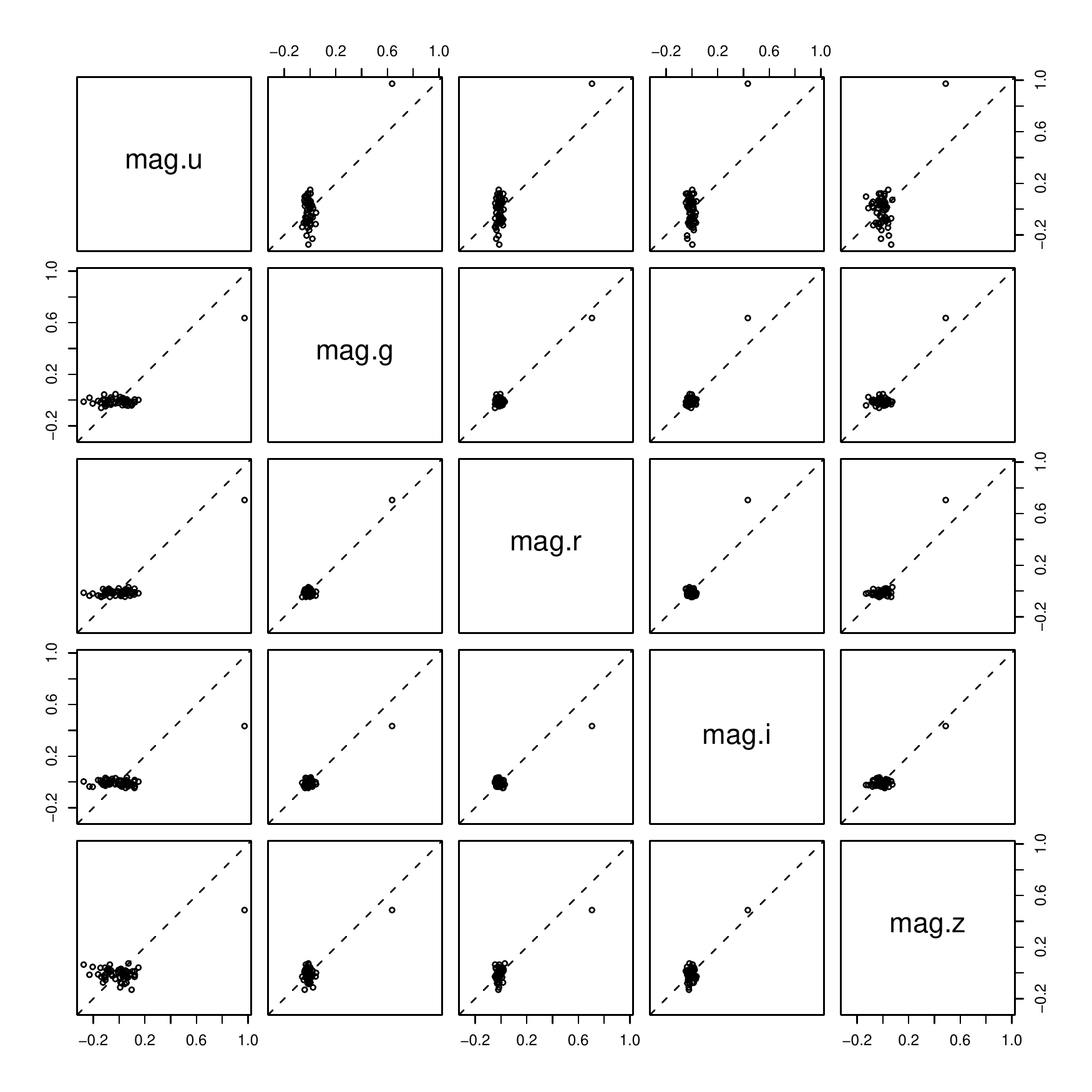}
\caption{Pairwise scatterplots of the centred magnitudes of a random star flagged variable from SDSS Stripe 82. The dashed line is a visual aid to assess eventual elongation of the point cloud.}
\label{fig:outlier}       
\end{figure} 

The application is usually not so straightforward as Fig.Ê\ref{fig:pca} suggests. Fig. \ref{fig:outlier} illustrates this with an SDSS point source. Observations made at different wavelengths often have different errors, as here the $u$-band and to a lesser extent, $z$. Since PCA turns into the direction of highest variance, in such a case it will indicate only the direction of the noisiest band. Therefore, we need to scale the observations to obtain unit variance in every band. Among the many options, the so-called variance stabilizing transformation proved to be the best, which, in addition, takes into account the dependence of errors on magnitudes, an effect that can be non-negligible for the faint bands and for large-amplitude variability. 

Another issue also observable in Fig. \ref{fig:outlier} is the presence of outliers. Obviously, they have a strong effect on the PCA fit: one outlier can turn the PC1 direction towards itself, falsifying completely the results. We dealt with this problem by applying a robust version  of PCA, the minimum covariance determinant method \cite{rousseeuw85}. Its tuning parameter is adjusted so that single outliers do not have strong effect on the fitted model, but a few consistently located distant points do distort the fit. The goal of this was to find a balance between two contradictory aims: decrease the effect of true erroneous data, but keep that of observations from strongly skewed light curves, most notably from eclipsing binaries.

The most important results of the principal component fit, the direction of the PC1, the variance in PC1 and the time series of the projections of the points to the PC1 direction (hereafter simply called PC1 time series) were then used for the analysis of the SDSS Stripe 82 data.


\section{Results}
\label{sec:results}

\subsection{Variability detection} %

In the framework of principal component analysis, variability detection on a scaled 5-dimensional cloud of $N$ points is equivalent to checking the statistical significance of additional variation as compared to a five-dimensional standard normal variable sample of size $N$. Though the variances of the principal components are 1 if all the original variables had unit variance (as is the case with our scaling), the first principal component of even a standard normal sample will show a variance higher than 1,  due to stochastic fluctuations and to the fact that the procedure is aimed at selecting the maximum spread direction. Its distribution can be simulated by generating standard normal samples of size $N$, and perform PCA on them. The null hypothesis of the scaled points being compatible with an uncorrelated, unit-variance normal noise is then checked by comparing the observed PC1 variance to the quantiles of the simulated distribution.

\begin{figure}[h]
\sidecaption
\includegraphics[scale=.64]{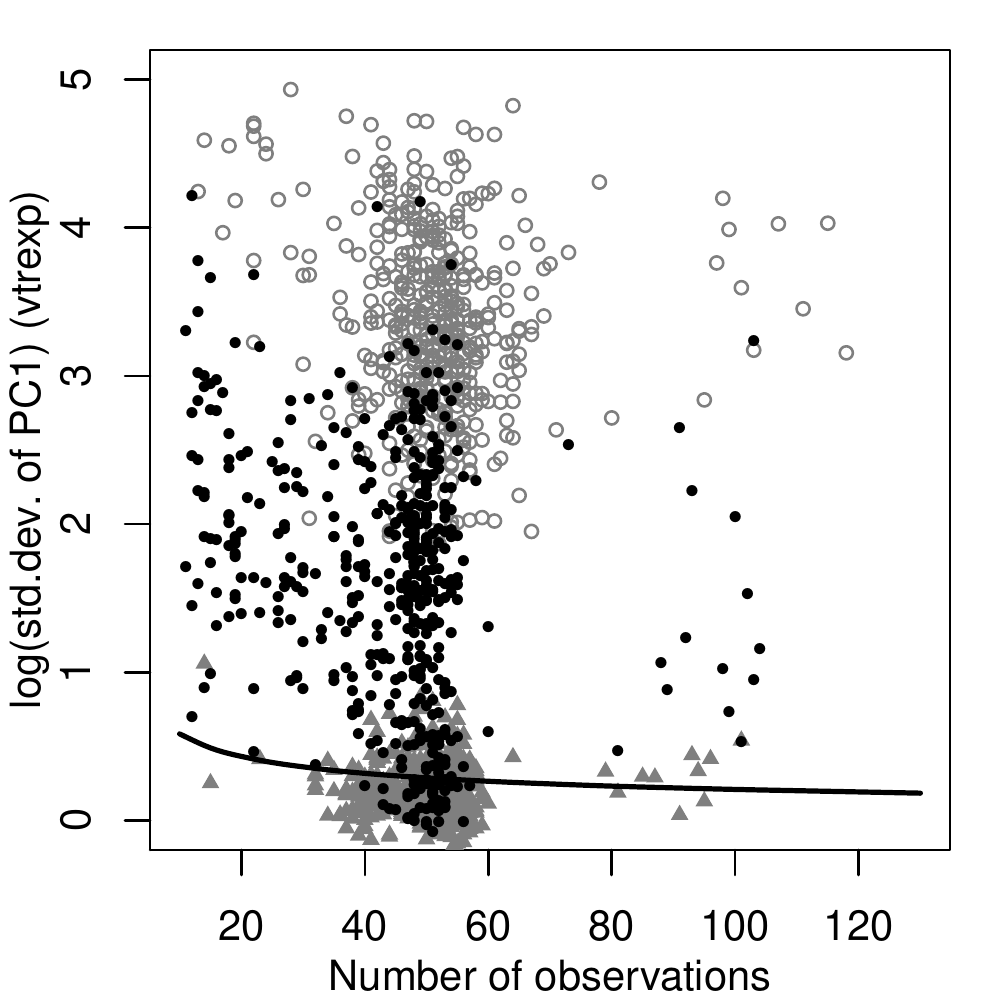}
\caption{The logarithm of the standard deviation of PC1 against the number of observations for the identified RR Lyrae sample (empty grey circles), for a random sample of objects flagged variable (black dots) and for a sample of objects flagged nonvariable (grey triangles). The 0.9999 quantile of the logarithm of the standard deviation of PC1 in simulated 5-variate standard normal samples is plotted as a black line.}
\label{fig:vardetect}       
\end{figure}

The logarithm of the estimated standard deviation of the first principal component for the  483 RR Lyrae-type stars of \cite{sesaretal10}, and a random sample of variable and non-variable stars of the same size from our data set is shown in Fig. \ref{fig:vardetect}, together with the 0.9999 quantile of the simulated distribution. The RR Lyraes are scattered in the highest regions of the plot, and thus are clearly identified as variable. The majority of the other variable objects are also detected as such, which shows that this criterion can yield a selection with similar completeness as the rms- and chi-squared statistics cuts in $g$ and $r$-band. The presence of objects flagged originally as non-variable above the 0.9999 quantile may be due to micro-variability or correlated errors that are detected by PCA, but missed by single-band analysis that is insensitive to correlations. 
Under the line, we can found several objects that were flagged as variable by the rms and chi-squared cuts but not spotted as variable by the PCA; one reason for this may be under-estimated errors, which caused the traditional methods over-estimate the intrinsic variability, but affected much less PCA.

\subsection{Period search} %

\begin{figure}[b]
\sidecaption[t]
\includegraphics[scale=.56]{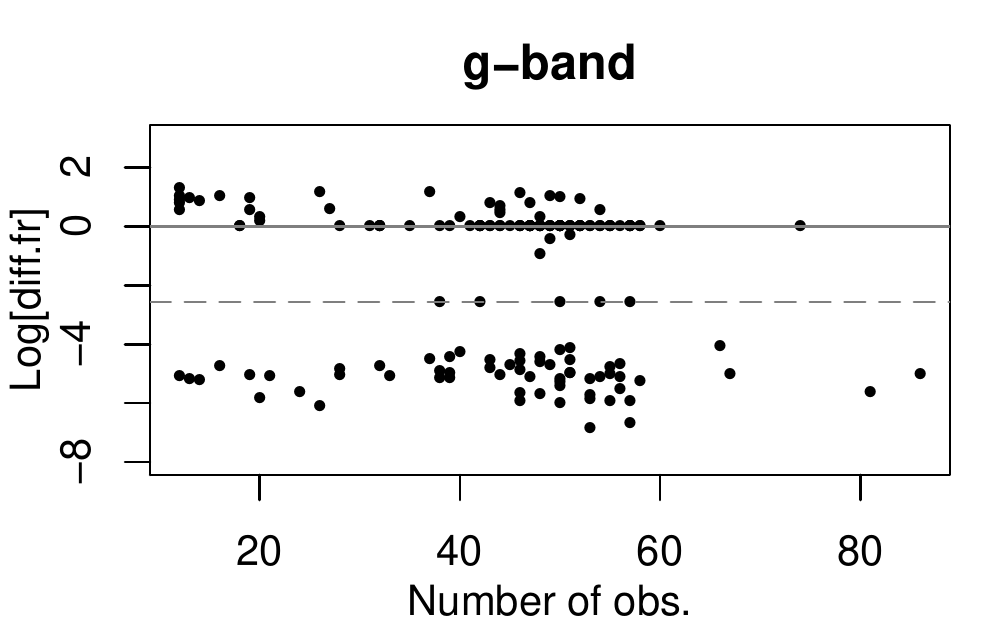}
\includegraphics[scale=.56]{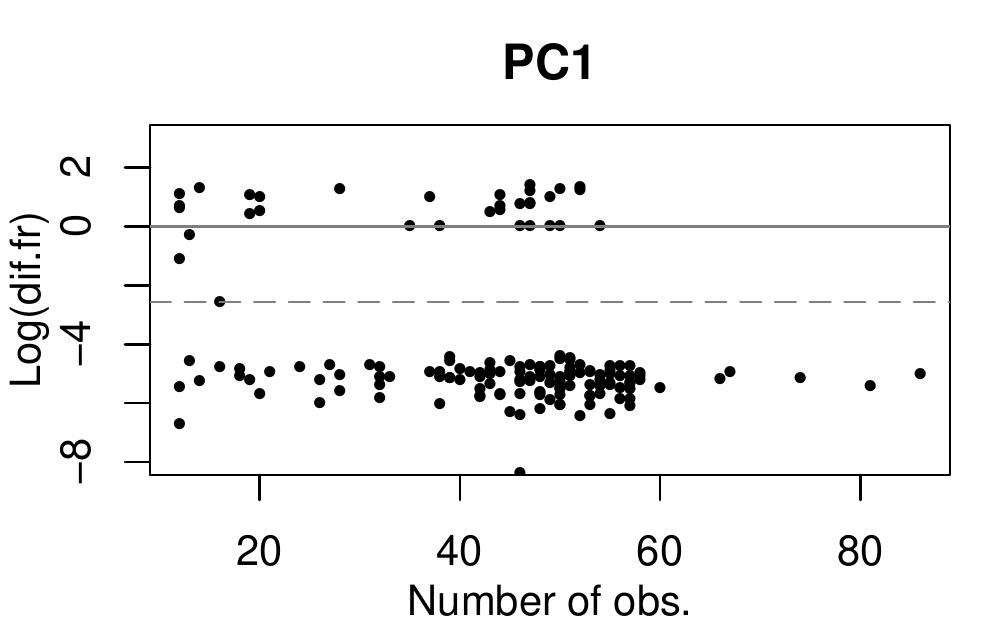}
\caption{The logarithm of the difference between the true and the found frequency in 1/day for sinusoidal light curve simulations with SDSS error distributions on $g$-band (left panel) and on PC1 (right panel). The solid line is the daily alias, the dashed line is the yearly alias.}
\label{fig:frdiff}       
\end{figure}

We applied the generalized least-squares method \cite{zechmeisterkurster09} to simulated data, first on $g$-band with the usual weighting based on the errors, then on the PC1 time series with a robust weighting based on truncated Mahalanobis distances. The simulations were constructed to imitate the bandwise error distributions of the SDSS, using sine waves with realistic (0-0.4 mag) amplitude sets and sampled at randomly selected real SDSS cadences. Fig. \ref{fig:frdiff} shows the results, the logarithm of the difference between the true and the found frequencies versus the number of observations. The advantages of the period search on the PC1 time series are clear: while using the $g$-band, many daily and yearly aliases were found instead of the true frequency (points on the solid and dashed lines), most of these disappeared when performing the search on the PC1 time series. Also, period search is slightly more successful for sparsely sampled time series on PC1 than on $g$-band.  

\subsection{Classification} %

\begin{figure}[h]
\sidecaption
\includegraphics[scale=.45]{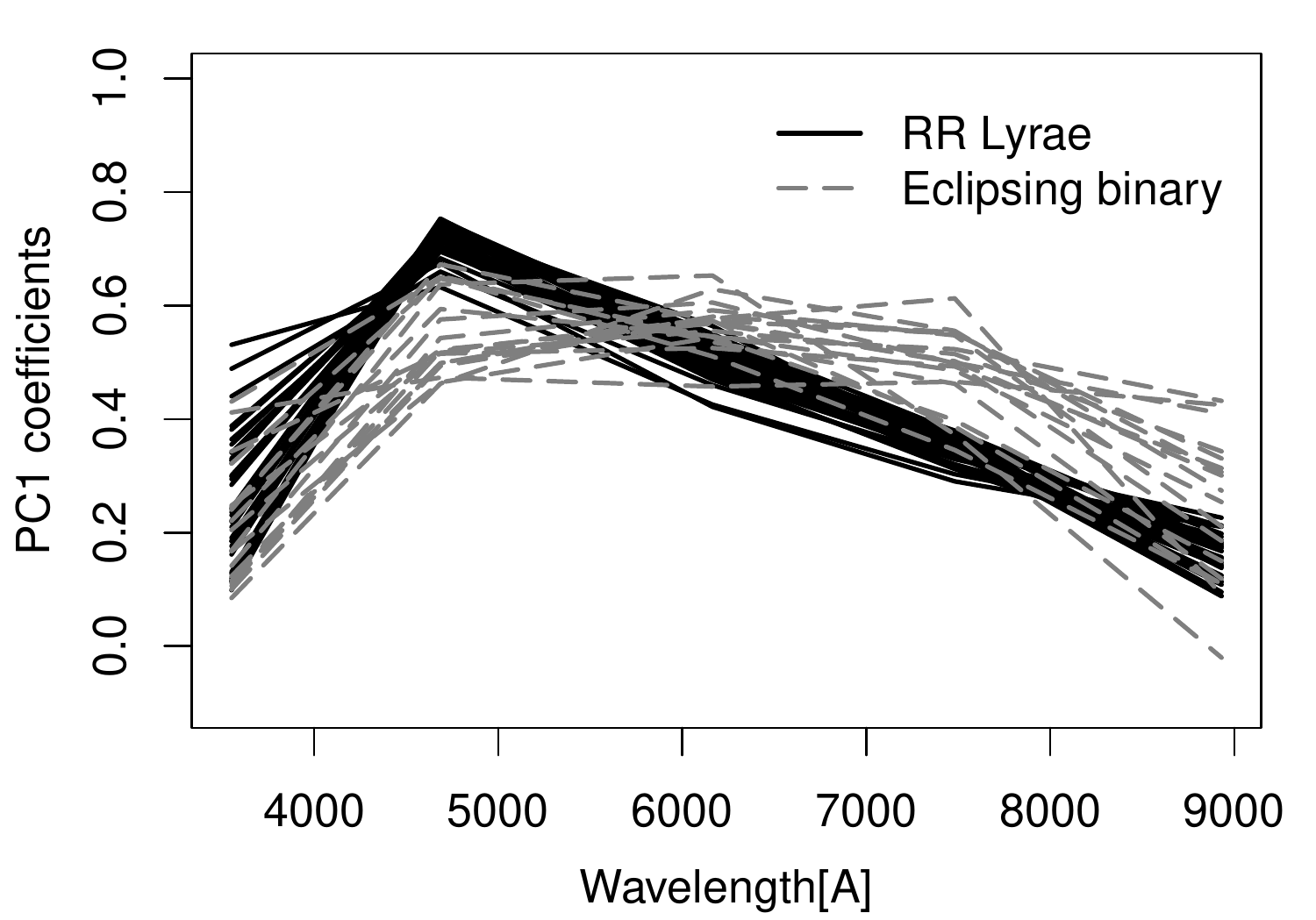}
\caption{PC1 spectra, i.e. PC1 coefficients of the various bands for the identified RR Lyraes of \cite{sesaretal10} (solid black) and a visually selected eclipsing binary candidate set (dashed grey).}
\label{fig:PC1sp} 
\end{figure}

Automated classification of variable stars uses attributes like period, amplitude, harmonic amplitudes and phases, colors, skewness and kurtosis of the observations. Many of these quantities can be defined for the PC1 time series too, and we can add to the attribute list the coefficients of the linear combination producing the PC1. If the average error of the bands were similar, the PC1 coefficients would give the highest weight to the band with the largest variability amplitude. Therefore, the pattern would be different for a pulsating variable and for an eclipsing binary: the first has the largest variability in the blue bands, and a corresponding PC1 spectrum peaky at the blue wavelengths, whereas an eclipsing binary with components of similar masses and similar colors will show an approximately horizontal shape. Scaling with the different average error sizes of the bands results in a PC1 spectrum distorted in a way characteristic to the survey. For SDSS errors, this appears in the small coefficients on the $u$ and the $z$ bands, where errors are larger than the other bands.  Fig. \ref{fig:PC1sp} illustrates the effect. The black lines show known RR Lyrae stars from \cite{sesaretal10}, the dashed grey lines are PC1 spectra of visually selected eclipsing binaries.  The small coefficients of the noisy $u$ and $z$ bands appear clearly on the profiles, but we can observe the strong peak on the $g$-band for the pulsating RR Lyraes and the flat-topped shape of the eclipsing binary sample. The discernible difference may be efficient help in the distinction between pulsating variables with symmetrical light curve and EW-type eclipsing binaries. 

We tested whether PCA methods are able to detect further RR Lyrae variables unidentified so far in the Stripe 82 sample, and to select another interesting class of variables, the SX Phoenicis type stars. These are metal-poor Population II radial-mode pulsating objects with light curves similar to that of RR Lyraes. They too obey a period-luminosity relationship \cite{mcnamara97,mcnamaraetal07}, and therefore can be used to map the old structures of the Galaxy. Using the traditional and the novel attributes, we constructed a training set by iterating  visual inspection and automated Random Forest \cite{breiman} selection, composed of three types of objects: first, a candidate SX Phoenicis sample of around 90 stars, showing pulsational-type PC1 spectrum, characteristic light curve shape, short period $\lesssim 0.1$ day, and located in the appropriate region of the $u-g$, $g-r$ color-color diagram; second, the known RR Lyrae sample of \cite{sesaretal10}; and third, a large mixed set of all other stars. 
In addition, we selected a number of other SX Phe and over a hundred new RR Lyrae candidates. For the majority of the new candidates spectroscopic data and metallicity estimates are not available, so confirming  the type of SX Phe candidate set is still future work. The RR Lyrae candidates are under study. The training set will be used in the future to classify a broader Stripe 82 data set containing also faint objects.


\section{Summary}
\label{sec:summary}

We tested principal component analysis as a way to combine quasi-simultaneous multi-band time series into one time series to obtain better signal-to-noise ratio and to extract a summary information about correlated variations on the SDSS Stripe 82 point sources. We derived a variability detection criterion that bases the decision on the existence of cross-band correlations, achieves improvement in period search results over single-band analysis, and formulated new, useful attributes for classification. The methods produced a promising set of new candidate RR Lyrae and SX Phoenicis variables, both of which can be used to trace halo structures in the Galaxy. We continue the work towards the extension of the methods to faint objects.


\begin{thebibliography}{99}

\bibitem{adelmanmccarthyetal08} J.~K.~Adelman-McCarthy \emph{et al.} (2008), ApJS {\bf 175}, p. 297


\bibitem{bramichetal08} D.~M.~Bramich \emph{et al.} (2008), MNRAS {\bf 386}, p. 77

\bibitem{breiman} L.~Breiman (2001), Machine Learning {\bf 45}, p. 5

\bibitem{friemanetal08} J.~A.~Frieman \emph{et al.} (2008), AJ {\bf 135}, p. 338

\bibitem{gaiaweb} http://www.rssd.esa.int/index.php?project=GAIA\&page=index

\bibitem{jolliffe02} I.~T.~Jolliffe, \emph{Principal Component Analysis}, 2nd edn. (Springer, New York, 2002)

\bibitem{hastieetal09} T.~Hastie, R.~Tibshirani, J.~Friedman, \emph{The Elements of Statistical Learning}, 2nd edn. (Springer Science+Business Media, 2009)

\bibitem{mcnamara97} D.~H.~McNamara (1997), PASP {\bf 109}, p. 1221 

\bibitem{mcnamaraetal07} D.~H.~McNamara, G.~Clementini, M.~Marconi (2007), AJ {\bf 133}, p. 2752

\bibitem{rousseeuw85} P.~J.~Rousseeuw, in \emph{Mathematical Statistics and Applications Vol. B}, eds. W.~Grossmann, G.~Pflug, I.~Vincze and W.Wertz, Dordrecht: Reidel (1985), p. 28

\bibitem{sdssweb} http://www.sdss.org/

\bibitem{sesaretal07} B.~Sesar \emph{et al.} (2007), AJ {\bf 134}, p. 2236

\bibitem{sesaretal10} B.~Sesar \emph{et al.} (2010), ApJ {\bf 708}, p. 717


\bibitem{zechmeisterkurster09} M.~Zechmeister, M.~K{\"u}rster (2009), A\&A {\bf 496}, p. 577

\end{thebibliography}
\end{document}